\documentclass[preprintnumbers,amsmath,amssymbm,prd]{revtex4}
\usepackage{epsfig}
\usepackage{graphicx}

\begin{document}
\title{Stationary resonances of rapidly-rotating Kerr black holes}
\author{Shahar Hod}
\affiliation{The Ruppin Academic Center, Emeq Hefer 40250, Israel}
\affiliation{ } \affiliation{The Hadassah Institute, Jerusalem
91010, Israel}
\date{\today}

\begin{abstract}
\ \ \ The Klein-Gordon equation for a massive scalar field in the
background of a rapidly-rotating Kerr black hole is studied {\it
analytically}. In particular, we derive a simple formula for the
stationary (marginally-stable) resonances of the field in the
black-hole spacetime. The analytically derived formula is shown to
agree with direct numerical computations of the resonances. Our
results provide an upper bound on the instability regime of
rapidly-rotating Kerr black holes to massive scalar perturbations.
\end{abstract}
\bigskip
\maketitle


\section{Introduction}

The `no-hair' conjecture \cite{Whee,Car}, put forward by Wheeler
more than four decades ago, asserts that stationary black-hole
spacetimes should be described by the Kerr-Newman metric. This
conjecture therefore suggests that stationary black holes can be
characterized by only three externally observable parameters: mass,
charge, and angular momentum.

According to the no-hair conjecture, it is expected that static
fields (with the exception of the electric field which is associated
with a globally conserved charge) cannot survive in the exterior of
black holes \cite{Whee,Car,BekTod2,Nun,Hod11,Chas,BekVec,Hart}. In
particular, such fields are expected to be radiated away to infinity
or to be swallowed by the black hole itself \cite{BekTod2,Hod11}.
Massless test fields indeed follow this scenario: their relaxation
phase in the exterior of black holes is characterized by
`quasinormal ringing', damped oscillations with a discrete spectrum
\cite{Nollert1,Ber1} (see also \cite{QNMlist} and references
therein). These characteristic oscillations are then followed by
late-time decaying tails \cite{Tails1,Tails2}.

However, it turns out that (non-static) {\it massive} scalar fields
\cite{Notebos} can survive in the exterior of {\it rotating} black
holes due to the well-known phenomena of superradiant scattering
\cite{Zel,PressTeu1,PressTeu2,Dam,Zour,Det,Furu,CarDias,Dolan,HodHod,Beyer2,HodN,Dolan2}:
a bosonic field of the form $e^{im\phi}e^{-i\omega t}$ impinging on
a rotating Kerr black hole can be amplified as it scatters off the
hole if it satisfies the superradiant condition
\begin{equation}\label{Eq1}
\omega<m\Omega\  ,
\end{equation}
where
\begin{equation}\label{Eq2}
\Omega={{a}\over{2Mr_+}}\
\end{equation}
is the angular velocity of the black-hole horizon. Here $M, Ma$, and
$r_+$ are the black-hole mass, angular momentum, and horizon-radius,
respectively. If in addition the scalar field has a non-zero rest
mass, then the mass term (the gravitational attraction between the
black hole and the massive field) effectively works as a mirror,
preventing the field from escaping to infinity.

In a seminal work, Detweiler \cite{Det} studied the Klein-Gordon
equation for the black-hole-scalar-field system in the regime
$M\mu\ll1$, that is in the regime where the Compton wavelength of
the field is much larger than the length-scale set by the black
hole. (Here $\mu\equiv {\cal M}G/\hbar c$, where ${\cal M}$ is the
mass of the field. We shall use natural units in which $G=c=1$
\cite{Noteunit}.) Using Eqs. $(18)$ and $(26)$ of \cite{Det} one
finds that marginally stable modes (that is, {\it stationary} modes
which are characterized by $\Im\omega=0$) of the massive scalar
field exist for the marginal frequency
\begin{equation}\label{Eq3}
\omega=m\Omega\
\end{equation}
with the discrete spectrum
\begin{equation}\label{Eq4}
\mu=m\Omega\Big[1+{1\over
2}\Big({{mM\Omega}\over{l+1+n}}\Big)^2+O[(M\Omega)^4]\Big]\
\end{equation}
of the field-masses. Here $l$ is the spherical harmonic index of the
mode, $m$ is the azimuthal harmonic index with $-l\leq m\leq l$, and
$n$ is the resonance parameter which is a non-negative integer
\cite{Det}.

It should be emphasized that the formula (\ref{Eq4}) is only valid
in the regime $M\mu\ll1$ studied in \cite{Det}. Thus, the formula
(\ref{Eq4}) for the field-masses of the stationary resonances is
only valid for {\it slowly}-rotating black holes (that is, in the
regime $M\Omega\ll1$).

The main goal of the present study is to obtain an {\it analytical}
formula for the field-masses of the stationary resonances in the
regime of {\it rapidly}-rotating black holes with $a\approx M$ (that
is, for $M\Omega\simeq {1/2})$. It is worth mentioning that we have
recently obtained a simple upper bound on the field-masses of the
stationary (marginally-stable) resonances \cite{HodN}:
\begin{equation}\label{Eq5}
\mu<\sqrt{2}m\Omega\  .
\end{equation}
This upper bound is valid in the entire range $0\leq a/M\leq 1$ of
the dimensionless black-hole spin. Note that the formula (\ref{Eq4})
above (which is only valid in the $a\ll M$ regime) conforms to this
upper bound.

\section{Description of the system}

The physical system we consider consists of a test scalar field
$\Psi$ coupled to a rotating Kerr black hole of mass $M$ and
angular-momentum per unit mass $a$. In Boyer-Lindquist coordinates
$(t,r,\theta,\phi)$ the spacetime metric is given by
\cite{Chan,Kerr}
\begin{eqnarray}\label{Eq6}
ds^2=-\Big(1-{{2Mr}\over{\rho^2}}\Big)dt^2-{{4Mar\sin^2\theta}\over{\rho^2}}dt
d\phi+{{\rho^2}\over{\Delta}}dr^2
+\rho^2d\theta^2+\Big(r^2+a^2+{{2Ma^2r\sin^2\theta}\over{\rho^2}}\Big)\sin^2\theta
d\phi^2,
\end{eqnarray}
where $\Delta\equiv r^2-2Mr+a^2$ and $\rho\equiv
r^2+a^2\cos^2\theta$. The black-hole (event and inner) horizons are
located at the zeroes of $\Delta$:
\begin{equation}\label{Eq7}
r_{\pm}=M\pm(M^2-a^2)^{1/2}\  .
\end{equation}
We shall henceforth assume that the black hole is rapidly-rotating
(near-extremal) with $a\approx M$.

The dynamics of a massive scalar field $\Psi$ in the Kerr spacetime
is governed by the Klein-Gordon equation \cite{Teuk}
\begin{equation}\label{Eq8}
(\nabla^a \nabla_a -\mu^2)\Psi=0\  .
\end{equation}
One may decompose the field as
\begin{equation}\label{Eq9}
\Psi_{lm}(t,r,\theta,\phi)=e^{im\phi}S_{lm}(\theta;a\omega)R_{lm}(r;a\omega)e^{-i\omega
t}\ ,
\end{equation}
where $\omega$ is the (conserved) frequency of the mode. (We shall
henceforth omit the indices $l$ and $m$ for brevity.) With the
decomposition (\ref{Eq9}), $R$ and $S$ obey radial and angular
equations both of confluent Heun type coupled by a separation
constant $K(a\omega)$ \cite{Heun,Fiz1,Teuk,Abram,Stro,Hodasy}.

The angular functions $S(\theta;a\omega)$ are the spheroidal
harmonics which are solutions of the angular equation
\cite{Heun,Fiz1,Teuk,Abram,Stro,Hodasy}
\begin{eqnarray}\label{Eq10}
{1\over {\sin\theta}}{\partial \over
{\partial\theta}}\Big(\sin\theta {{\partial
S}\over{\partial\theta}}\Big) +\Big[K+a^2(\mu^2-\omega^2)
-a^2(\mu^2-\omega^2)\cos^2\theta-{{m^2}\over{\sin^2\theta}}\Big]S=0\
.
\end{eqnarray}
The angular functions are required to be regular at the poles
$\theta=0$ and $\theta=\pi$. These boundary conditions pick out a
discrete set of eigenvalues $\{K_{lm}\}$ labeled by the integers $l$
and $m$. For $a^2(\mu^2-\omega^2)\lesssim m^2$ one can treat
$a^2(\omega^2-\mu^2)\cos^2\theta$ in Eq. (\ref{Eq10}) as a
perturbation term on the generalized Legendre equation and obtain
the perturbation expansion \cite{Abram}
\begin{equation}\label{Eq11}
K_{lm}+a^2(\mu^2-\omega^2)=l(l+1)+\sum_{k=1}^{\infty}c_ka^{2k}(\mu^2-\omega^2)^k\
\end{equation}
for the separation constants $K_{lm}$. The expansion coefficients
$\{c_k(l,m)\}$ are given in Ref. \cite{Abram}.

The radial Teukolsky equation is given by \cite{Teuk,Stro}
\begin{equation}\label{Eq12}
\Delta{{d} \over{dr}}\Big(\Delta{{dR}\over{dr}}\Big)+\Big[H^2
+\Delta[2ma\omega-K-\mu^2(r^2+a^2)]\Big]R=0\ ,
\end{equation}
where $H\equiv (r^2+a^2)\omega-am$. We are interested in solutions
of the radial equation (\ref{Eq12}) with the physical boundary
conditions of purely ingoing waves at the black-hole horizon (as
measured by a comoving observer) and a bounded (decaying) solution
at spatial infinity
\cite{Dam,Zour,Det,Furu,CarDias,Dolan,HodHod,Beyer2,HodN,Dolan2}.
That is,
\begin{equation}\label{Eq13}
R \sim
\begin{cases}
{1\over r}e^{-\sqrt{\mu^2-\omega^2}y} & \text{ as }
r\rightarrow\infty\ \ (y\rightarrow \infty)\ ; \\
e^{-i (\omega-m\Omega)y} & \text{ as } r\rightarrow r_H\ \
(y\rightarrow -\infty)\ ,
\end{cases}
\end{equation}
where the ``tortoise" radial coordinate $y$ is defined by
$dy=[(r^2+a^2)/\Delta]dr$.

Note that a bound state (a state decaying exponentially at spatial
infinity) is characterized by $\omega^2<\mu^2$. The boundary
conditions (\ref{Eq13}) single out a discrete set of complex
resonances $\{\omega_n(\mu)\}$ which correspond to the bound states
of the massive field
\cite{Dam,Zour,Det,Furu,CarDias,Dolan,HodHod,Beyer2,HodN,Dolan2,NoteWill,Will}.
The stationary (marginally-stable) resonances, which are the
solutions we are interested in in this paper, are characterized by
$\Im\omega=0$.

\section{The stationary scalar resonances}

As we shall now show, the field (\ref{Eq9}) with the marginal
frequency (\ref{Eq3}) describes a {\it stationary} resonance of the
Klein-Gordon equation (\ref{Eq8}) in the black-hole spacetime. In
particular, we shall now derive an analytical formula for the
discrete spectrum $\{M\mu(m,l,n)\}$ of field-masses which satisfy
the stationary resonance condition $\Im\omega=0$. To that end, it is
convenient to define new dimensionless variables
\begin{equation}\label{Eq14}
x\equiv {{r-r_+}\over {r_+}}\ \ ;\ \ \tau\equiv 8\pi
MT_{BH}={{r_+-r_-}\over {r_+}}\ \ ;\ \ k\equiv 2m\Omega r_+\ \ ;\ \
\epsilon\equiv \sqrt{\mu^2-(m\Omega)^2}r_+\ ,
\end{equation}
in terms of which the radial equation (\ref{Eq12}) becomes
\begin{equation}\label{Eq15}
x(x+\tau){{d^2R}\over{dx^2}}+(2x+\tau){{dR}\over{dx}}+VR=0\  ,
\end{equation}
where $V\equiv
H^2/r^2_+x(x+\tau)-K_{lm}+2m^2a\Omega-\mu^2[r^2_+(x+1)^2+a^2]$ and
$H=kr_+x({1\over 2}x+1)$.

We first consider the radial equation (\ref{Eq15}) in the far region
$x\gg \tau$. Then Eq. (\ref{Eq15}) is well approximated by
\begin{equation}\label{Eq16}
x^2{{d^2R}\over{dx^2}}+2x{{dR}\over{dx}}+V_{\text{far}}R=0\  ,
\end{equation}
where $V_{\text{{far}}}=-(\epsilon
x)^2+k^2x/2+[-K_{lm}+2m^2a\Omega+k^2-\mu^2(r^2_++a^2)]$. A solution
of Eq. (\ref{Eq16}) that satisfies the boundary condition
(\ref{Eq13}) can be expressed in terms of the confluent
hypergeometric functions $M(a,b,z)$ \cite{Morse,Abram}
\begin{equation}\label{Eq17}
R=C_1(2\epsilon)^{{1\over 2}+\beta}x^{-{1\over 2}+\beta}e^{-\epsilon
x}M({1\over 2}+\beta-\kappa,1+2\beta,2\epsilon x)+C_2(\beta\to
-\beta)\ ,
\end{equation}
where $C_1$ and $C_2$ are constants. Here
\begin{equation}\label{Eq18}
\beta^2\equiv K_{lm}+{1\over 4}+\mu^2(r^2_++a^2)-k^2-2m^2a\Omega\
\end{equation}
and
\begin{equation}\label{Eq19}
\kappa\equiv {{{1\over 4}k^2-\epsilon^2}\over{\epsilon}}\ .
\end{equation}
The notation $(\beta\to -\beta)$ means ``replace $\beta$ by $-\beta$
in the preceding term."

We next consider the near horizon region $x\ll 1$. The radial
equation is given by Eq. (\ref{Eq15}) with $V\to
V_{\text{near}}\equiv
-K_{lm}+2m^2a\Omega-\mu^2(r^2_++a^2)+k^2x/(x+\tau)$. The physical
solution obeying the ingoing boundary condition at the horizon is
given by \cite{Morse,Abram}
\begin{equation}\label{Eq20}
R=\Big({x\over \tau}+1\Big)^{-ik}{_2F_1}({1\over 2}+\beta-ik,{1\over
2}-\beta-ik;1;-x/\tau)\  ,
\end{equation}
where $_2F_1(a,b;c;z)$ is the hypergeometric function.

The solutions (\ref{Eq17}) and (\ref{Eq20}) can be matched in the
overlap region $\tau\ll x\ll 1$. The $x\ll 1$ limit of Eq.
(\ref{Eq17}) yields \cite{Morse,Abram}
\begin{equation}\label{Eq21}
R\to C_1(2\epsilon)^{{1\over 2}+\beta}x^{-{1\over
2}+\beta}+C_2(\beta\to -\beta)\  .
\end{equation}
The $x\gg \tau$ limit of Eq. (\ref{Eq20}) yields \cite{Morse,Abram}
\begin{equation}\label{Eq22}
R\to \tau^{{1\over 2}-\beta}{{\Gamma(2\beta)}\over{\Gamma({1\over
2}+\beta-ik)\Gamma({1\over 2}+\beta+ik)}}x^{-{1\over
2}+\beta}+(\beta\to -\beta)\  .
\end{equation}
By matching the two solutions in the overlap region one finds
\begin{equation}\label{Eq23}
C_1=\tau^{{1\over 2}-\beta}{{\Gamma(2\beta)}\over{\Gamma({1\over
2}+\beta-ik)\Gamma({1\over 2}+\beta+ik)}}(2\epsilon)^{-{1\over
2}-\beta}\ ,
\end{equation}
and
\begin{equation}\label{Eq24}
C_2=\tau^{{1\over 2}+\beta}{{\Gamma(-2\beta)}\over{\Gamma({1\over
2}-\beta-ik)\Gamma({1\over 2}-\beta+ik)}}(2\epsilon)^{-{1\over
2}+\beta}\ .
\end{equation}

Approximating Eq. (\ref{Eq17}) for $x\to\infty$ one gets
\cite{Morse,Abram}
\begin{eqnarray}\label{Eq25}
R&\to&
\Big[C_1(2\epsilon)^{-\kappa}{{\Gamma(1+2\beta)}\over{\Gamma({1\over
2}+\beta-\kappa)}}x^{-1-\kappa}+C_2(\beta\to -\beta)\Big]e^{\epsilon
x}\nonumber
\\&& + \Big[C_1(2\epsilon)^{\kappa}{{\Gamma(1+2\beta)}\over{\Gamma({1\over
2}+\beta+\kappa)}}x^{-1+\kappa}(-1)^{-{1\over
2}-\beta+\kappa}+C_2(\beta\to -\beta)\Big]e^{-\epsilon x}\ .
\end{eqnarray}
A bound state is characterized by a decaying field at spatial
infinity. The coefficient of the growing exponent $e^{\epsilon x}$
in Eq. (\ref{Eq25}) should therefore vanish. Taking cognizance of
Eqs. (\ref{Eq23})-(\ref{Eq25}), one finds the characteristic
equation
\begin{equation}\label{Eq26}
{1\over{\Gamma({1\over
2}+\beta-\kappa)}}=\Big[{{\Gamma(-2\beta)}\over{\Gamma(2\beta)}}\Big]^2{{\Gamma({1\over
2}+\beta-ik)\Gamma({1\over 2}+\beta+ik)}\over{\Gamma({1\over
2}-\beta-ik)\Gamma({1\over 2}-\beta+ik)\Gamma({1\over
2}-\beta-\kappa)}}\big(2\epsilon\tau\big)^{2\beta}\
\end{equation}
for the stationary bound states of the massive scalar field. Note
that the r.h.s. of Eq. (\ref{Eq26}) is of order $O(\tau^{2\beta})\ll
1$.
Thus, using the well-known pole structure of the Gamma functions
\cite{Abram}, one finds that the resonance condition (\ref{Eq26})
can be written as
\begin{equation}\label{Eq27}
{1\over 2}+\beta-\kappa=-n+O(\tau^{2\beta})\ ,
\end{equation}
where $n\geq 0$ is a non-negative integer.

We shall henceforth assume that $2\beta>1$ \cite{Notebeta} and
expand all quantities to first order in the small parameter $\tau$.
Taking cognizance of Eqs. (\ref{Eq11}), (\ref{Eq14}), (\ref{Eq18}),
and (\ref{Eq19}), one finds
\begin{equation}\label{Eq28}
\beta^2=(l+{1\over 2})^2-{3\over 2}m^2+{1\over
4}m^2\tau+O(\tau^2,\epsilon^2)\
\end{equation}
and
\begin{equation}\label{Eq29}
\kappa={{m^2}\over{4\epsilon}}-\epsilon+O(\tau^2)\  .
\end{equation}
Substituting Eqs. (\ref{Eq28})-(\ref{Eq29}) into (\ref{Eq27}), one
finds that the resonance condition for the stationary modes can be
expressed as a polynomial equation for the dimensionless variable
$\epsilon$:
\begin{equation}\label{Eq30}
4\big[(2l+1)^2-m^2(4-\tau)-(2n+1)^2\big]\epsilon^2+4m^2(2n+1)\epsilon-m^4+O(\tau^{2\alpha},\epsilon^3)=0\
,
\end{equation}
where $\alpha\equiv\min\{1,\beta\}$. This simple equation can easily
be solved to yield \cite{Notethird}
\begin{equation}\label{Eq31}
\bar\epsilon(l,m,n)\equiv{{\epsilon}\over{m}}={{m}\over{2(\ell+1+2n)}}
-{{m^3}\over{4\ell(\ell+1+2n)^2}}\tau+O(\tau^{2\alpha},{\bar\epsilon}^3)\
,
\end{equation}
where $\ell\equiv\sqrt{(2l+1)^2-4m^2}$ \cite{Noteepsi}. The
field-masses of the stationary resonances are given by
$\mu=\sqrt{(m\Omega)^2+(\epsilon/r_+)^2}$ [see Eq. (\ref{Eq14})],
which implies
\begin{equation}\label{Eq32}
\mu=m\Omega\big[1+2{\bar\epsilon}^2+O(\tau^2,{\bar\epsilon}^4)\big]\
.
\end{equation}

\section{Numerical confirmation}

We shall now test the accuracy of the analytically derived formula
(\ref{Eq32}) for the field-masses of the stationary resonances. The
stationary resonances can be computed using standard numerical
techniques, see \cite{Dolan,Dolan2} for details. In Table
\ref{Table1} we present a comparison between the {\it analytically}
derived field-masses of the stationary resonances (\ref{Eq32}), and
the {\it numerically} computed field-masses \cite{Dolan,Dolan2} for
the physically most interesting mode
\cite{Dam,Zour,Det,Furu,CarDias,Dolan,HodHod,Beyer2,HodN,Dolan2},
$l=m=1$ with $n=0$. We find an almost perfect agreement between the
two in the $\tau\ll 1$ ($a/M\gtrsim 0.99$) regime. In fact, one
finds that the agreement between the numerical data and the
analytical formula (\ref{Eq32}) is quite good already at $a/M=0.9$.
This is quite surprising since the assumption $\tau\ll 1$ breaks
down for this value of the dimensionless spin parameter.

\begin{table}[htbp]
\centering
\begin{tabular}{|c|c|c|c|c|c|c|}
\hline
\ \ $a/M$\ \ & \ 0.9\ \ & \ \ 0.95\ \ & \ \ 0.99\ \ & \ \ 0.995\ \ & \ \ 0.999\ \ & \ 1.0\ \ \\
\hline \ \ $\mu_{\text{ana}}/\mu_{\text{num}}$\ \ & \ \ 1.029\ \ \ &
\ \ 1.024\ \ \ & \ \ 1.007\ \ \ &
\ \ 1.006\ \ \ & \ \ 1.004\ \ \ & \ \ 1.003\ \ \ \\
\hline
\end{tabular}
\caption{Stationary resonances of a massive scalar field in the
background of a rapidly-rotating Kerr black hole. The data shown is
for the fundamental mode $l=m=1$ with $n=0$, see also
\cite{Dolan,Dolan2}. We display the ratio between the analytically
derived field-mass, $\mu_{\text{ana}}$, and the numerically computed
values, $\mu_{\text{num}}$. The agreement between the numerical data
and the analytical formula (\ref{Eq32}) is better than $3\%$ in the
$a/M\gtrsim 0.9$ regime. (In fact, the agreement becomes much better
than $1\%$ in the $a/M\gtrsim 0.99$ regime).} \label{Table1}
\end{table}

\section{Summary}

In summary, we have analyzed the spectrum of stationary
(marginally-stable) resonances of massive scalar fields in the
spacetime of rapidly-rotating (near-extremal) Kerr black holes. In
particular, we have obtained an analytical expression for the
field-masses of the stationary ($\Im\omega=0$) resonances, see Eqs.
(\ref{Eq31})-(\ref{Eq32}). We have shown that the analytically
derived formula agrees with direct numerical computations of the
resonances. 
Finally, we note that our
results [compare Eqs. (\ref{Eq5}) and (\ref{Eq32})] provide an
improved upper bound on the instability regime of rapidly-rotating
Kerr black holes to massive scalar perturbations.

\bigskip
\noindent
{\bf ACKNOWLEDGMENTS}
\bigskip

This research is supported by the Carmel Science Foundation. I thank
Yael Oren, Arbel M. Ongo and Ayelet B. Lata for stimulating
discussions.



\begin{thebibliography}{99}
\bibitem{Noteunit} In these units $\mu$ has the dimensions of 1/length.
\bibitem{NoteWill} We note that, in addition to the bound states of the massive
field, the field also has an infinite set of discrete quasinormal
resonances \cite{Will} which are characterized by outgoing waves at
spatial infinity.
\bibitem{Notebeta} In the $\tau\ll 1$ regime the condition $2\beta>1$ is satisfied by modes
with $m<\sqrt{2l(l+1)/3}$, see Eq. (\ref{Eq28}) below. In
particular, it is satisfied by the fundamental $l=m=1$ mode.
\bibitem{Notethird} We denote by $\epsilon$ the solution of the
quadratic equation (\ref{Eq30}). The $O(\epsilon^3)$ term in
(\ref{Eq30}) is $-16(2n+1)\epsilon^3$. In the $\tau\to 0$ limit this
term introduces a small correction of the form: $\epsilon\to
\epsilon(1+\delta)$ with
$\delta\simeq{{m^2(2n+1)}\over{\ell(\ell+1+2n)^2}}$. Note that in
the regime $2\beta>1$ that we consider ($m<\sqrt{2l(l+1)/3}$, see
\cite{Notebeta}), one has $\delta<{{{2\over
3}l(l+1)}\over{\sqrt{{4\over 3}l(l+1)+1}[\sqrt{{4\over
3}l(l+1)+1}+1]^2}}\ll 1$.
\bibitem{Noteepsi} Note that ${\bar\epsilon}^2\ll 1$ in the regime $2\beta>1$ that we consider ($m<\sqrt{2l(l+1)/3}$, see
\cite{Notebeta}). This justifies the perturbation expansion
(\ref{Eq11}).
\end{thebibliography}
\end{document}